\documentstyle[12pt]{article} 
\begin{document} 

\begin{center}
{\bf Domain Defects in Systems of Two Real Scalar Fields}\footnote{This 
work is supported in part by funds provided by the U.S. Department of 
Energy (D.O.E.) under cooperative research agreement \#DF-FC02-94ER40818.}
\end{center}

\begin{center}
D. Bazeia,$^{1,2}$ H. Boschi-Filho,$^{1,3}$ and F. A. Brito$^{2,4}$
\end{center}

\begin{center}
$^1$Center for Theoretical Physics\\
Laboratory for Nuclear Science and Department of Physics\\
Massachusetts Institute of Technology, Cambridge, Massachusetts 02139-4307
\end{center}

\begin{center}
$^2$Departamento de F\'\i sica, Universidade Federal da Para\'\i ba\\
Caixa Postal 5008, 58051-970 Jo\~ao Pessoa, Para\'\i ba, Brazil
\end{center}

\begin{center}
$^3$Instituto de F\'\i sica, Universidade Federal do Rio de Janeiro\\
Caixa Postal 68528, 21945-970 Rio de Janeiro, Rio de Janeiro, Brazil
\end{center}

\begin{center}
$^4$Departamento de F\'\i sica, Universidade Federal de Pernambuco\\
50670-901 Recife, Pernambuco, Brazil
\end{center}

\vskip .7cm

\begin{center}
MIT-CTP-2797
\end{center}

\vskip .7cm
 
\begin{center}
Abstract
\end{center}

In this work we investigate the role of the symmetry of the Lagrangian 
on the existence of defects in systems of coupled scalar fields. We focus
attention mainly on solutions where defects may nest defects. When
space is non-compact we find topological BPS and non-BPS solutions that
present internal structure. When space is compact the solutions are
nontopological sphalerons, which may be nested inside the topological defects.
We address the question of classical stability of these topological and
nontopological solutions and investigate how the thermal corrections may
modify the classical scenario.

\vskip .5in

PACS number(s): 11.10.Lm, 11.27.+d, 98.80.Cq 
 
\newpage 
 
\section{Introduction} 
\label{sec:intro}

In recent years much investigation on topological defects have been done 
with applications on diverse fields as for instance cosmology and 
particle physics \cite{vsh94,hki95}. In the present paper we are interested in
systems of coupled scalar fields where solutions like defects inside defects
may appear. Such situation was first investigated in the work of Witten
\cite{wit85}, within the context of superconducting strings -- see also the
work of Lazarides and Shafi \cite{lsh85} and of MacKenzie \cite{mac88}.
Several aspects of the possibility of topological defects being nested
inside topological defects are considered in
Refs.~{\cite{mor94,brs96,bba97,mor98,etb98}, and in 
Refs.~{\cite{baz95,mba96,cso97,shi97,shi98}} other related issues are also
investigated.

To introduce defects inside defects one needs to consider systems of at least
two fields, one of the fields being able to describe a specific defect, 
which will respond for nesting the other defect in its interior. The 
system of two fields should then be able to accommodate some defect and, 
in the core of this defect, be reduced to a system of the second field 
that is still able to generate the other defect. Since these defects are 
usually topological, and since topological defects appear in systems 
presenting spontaneous symmetry breaking, the complete system usually 
presents $G_1\times G_2$ symmetry, for instance $U(1)\times U(1)$ in the 
case of the superconducting strings examined by Witten \cite{wit85} or 
$Z_2\times Z_2$ domain walls, as considered for instance in the recent 
works \cite{brs96,bba97,mor98,etb98}. 

Although systems of two coupled fields engendering $G_1\times G_2$ symmetry
are known to support defects inside defects, we may wonder if defects with
internal structures appear after changing the symmetry of the system. Stated
differently, we may ask how the picture of defects inside defects modifies
when one changes the symmetry of the underlying Lagrangian, enlarging or
reducing the $G_1\times G_2$ group. This issue is of general interest and
may find direct applications in cosmology and in condensed matter. In the
present work we address this and other related questions in systems of two
real scalar fields, described via the Lagrangian density 
\begin{equation}\label{L} 
{\cal L}={1\over 2}\partial_\alpha \phi \partial^\alpha \phi 
+{1\over 2} \partial_\alpha\chi\partial^\alpha \chi-U(\phi,\chi)~.
\end{equation}
Although we are working in $3+1$ space-time dimensions we can search for
static solutions in the form $\phi=\phi(x)$ and $\chi=\chi(x)$, depending on
only one $(x^1=x)$ of the three spatial coordinates. In this case the
equations of motion are given by
\begin{eqnarray}
\label{eqmtx1} 
\frac{d^2\phi}{dx^2}&=&{\partial U\over\partial \phi}~,
\\
\label{eqmtx2} 
\frac{d^2\chi}{dx^2}&=&{\partial U\over\partial \chi}~.
\end{eqnarray}
We are interested in a special class of systems, in which the potential  
is determined by 
\begin{equation} 
\label{potential} 
U(\phi,\chi) 
=\frac{1}{2}\left(\frac{\partial H}{\partial\phi}\right)^2+ 
\frac{1}{2}\left(\frac{\partial H}{\partial\chi}\right)^2~.
\end{equation}
It is defined in terms of a smooth but otherwise arbitrary function
$H=H(\phi,\chi)$ and now the equations of motion become 
\begin{eqnarray} 
\label{eqmx1} 
{d^2\phi\over dx^2}&=&H_\phi H_{\phi\phi}+H_\chi H_{\chi\phi}~,\\  
\label{eqmx2} 
{d^2\chi\over dx^2}&=&H_\phi H_{\phi\chi}+H_\chi H_{\chi\chi}~, 
\end{eqnarray} 
where $H_\phi$ stands for $\partial H/\partial\phi$, and so forth.
The advantage of considering such potential is that the energy of static
configurations is minimized to the value
\begin{equation}
E_B=|H[\phi(+\infty),
\chi(+\infty)]-H[\phi(-\infty),\chi(-\infty)]|~,
\end{equation}
which depends just on the asymptotic values of the field configurations.
Furthermore, the second-order differential equations of motion are now
solved by field configurations that obey the first-order equations
\begin{eqnarray} 
\label{foeq1} 
{d\phi\over dx}&=&H_\phi~,\\  
\label{foeq2} 
{d\chi\over dx}&=&H_\chi~. 
\end{eqnarray} 
This result can be shown \cite{baz95} to follow the idea of
Bogomol'nyi, Prasad and Sommerfield (BPS) \cite{bps75} in the case of
coupled real scalar fields. This means that the potential in
Eq.~(\ref{potential}) can be considered as describing the bosonic sector of 
a larger supersymmetric theory, so that supersymmetric extensions of the 
Lagrangian (\ref{L}) can be readly constructed, as was considered for 
instance in Ref.~\cite{etb98,mba96} as well as in several other
investigations \cite{cso97,shi97,shi98}.

The present work is organized as follows. In the next
Sec.~{\ref{sec:solutions}} we search for defects inside defects
starting with a general potential. There we show that
defects may engender internal structure only when the system presents
parity symmetry of the type $Z_2\times Z_2$. Once this is
stablished we write down explicit solutions, examining both the second-order
equations of motion and the first-order equations that give BPS solutions.
We also consider the possibility of dealing with compact space, to
obtain periodic solutions similar to the sphaleron
solutions discussed by Manton and Samols \cite{msa88}. Such solutions are
known to be useful for discussing violation of the barionic number, and may
contribute to applications in cosmology. In
Sec.~{\ref{sec:stability}} we investigate classical or linear stability
of the solutions introduced in Sec.~{\ref{sec:solutions}}. Although some
of these solutions were already discussed in Refs.~{\cite{bba97,mor98}},
a complete investigation of their stability is still missing
and will be done in this work. In Sec.~{\ref{sec:temperature}}
we introduce finite temperature contributions to investigate
how the thermal effects may change the classical scenario. We end this paper
in Sec.~{\ref{sec:comments}}, where we present comments and conclusions.

\section{Symmetry and defects inside defects}
\label{sec:solutions}

We start with the simplest case in which a theory presents a $Z_2$ 
parity symmetry with nontrivial vacuum states for a single field $\phi$.
For
\begin{eqnarray} 
\label{U(phi)}
U(\phi)=\frac{1}{2}\,H_{\phi}^2~,
\end{eqnarray}
the standard $\phi^4$ model is obtained with the function
\begin{eqnarray} 
\label{H(phi)} 
H(\phi)=\lambda(\frac{1}{3}\phi^3-a^2\phi). 
\end{eqnarray}
A natural generalization for two coupled fields, in a renormalizable 
theory in (3+1) dimensions can be written as 
\begin{equation} 
\label{H_gen} 
H(\phi,\chi)=\lambda(\frac{1}{3}\phi^3-a^2\phi) 
+\mu\phi\chi^2+\nu\phi^2\chi+ 
\sigma(\frac{1}{3}\chi^3-b^2\chi), 
\end{equation} 
where $\lambda$, $\mu$, $\nu$, $\sigma$, $a$ and $b$ are real parameters.
Here we have the general potential 
\begin{eqnarray} 
\label{U_gen} 
U(\phi,\chi) 
&=&\frac{1}{2}(\lambda^2+\nu^2)\phi^4 
+2\nu(\lambda+\mu)\phi^3\chi 
+(\lambda\mu+2\nu^2+\sigma\nu+2\mu^2)\phi^2\chi^2\ 
\nonumber\\ 
&+&2\mu(\nu+\sigma)\phi\chi^3+ 
\frac{1}{2}(\mu^2+\sigma^2)\chi^4 
-(\lambda^2a^2+\sigma b^2\nu)\phi^2 
-(\lambda a^2\mu+\sigma^2 b^2)\chi^2 
\nonumber\\ 
&-&2(\lambda a^2\nu+\sigma b^2\mu)\phi\chi+\frac{1}{2} 
(\lambda^2 a^4+\sigma^2 b^4)~. 
\end{eqnarray} 
We now impose that the theory presents $Z_2\times Z_2$ symmetry. To ensure
invariance of the Lagrangian under the independent transformations
$\phi\rightarrow-\phi$ and $\chi\rightarrow-\chi$ we set 
\begin{eqnarray} 
\nu(\lambda+\mu)=0,\label{cond1} \\ 
\mu(\nu+\sigma)=0, \label{cond2} \\ 
\lambda\nu a^2+\sigma\mu b^2=0. \label{cond3} 
\end{eqnarray} 
There are four possibilities of satisfying the above
conditions; they are: 
\begin{eqnarray} 
(i) && \qquad \nu=\sigma=0, \\ 
(ii) && \qquad \lambda=\mu=0,\\ 
(iii) && \qquad \nu=\mu=0,\\ 
(iv) && \qquad \lambda=-\mu; \qquad \nu=-\sigma; \qquad a^2=- b^2. 
\end{eqnarray} 
The first possibility gives a model for which the function $H(\phi,\chi)$
reads 
\begin{eqnarray} 
\label{H_1} 
H_1(\phi,\chi)=\lambda(\frac{1}{3}\phi^3-a^2\phi)+\mu\phi\chi^2, 
\end{eqnarray} 
which was discussed recently in the literature \cite{brs96,bba97}. This is a
tipical example, shown to present topological solutions of the type of
defects inside defects. The second possibility leads to a model that is
described by 
\begin{eqnarray}\label{H_2} 
H_2(\phi,\chi)=\sigma(\frac{1}{3}\chi^3-b^2\chi)+\nu\phi^2\chi, 
\end{eqnarray} 
which is equivalent to the previous one,  
as can be seen by simply exchanging the fields $\phi$ and $\chi$ and 
redefining the parameters $\sigma$, $\nu$ and $b$.  
The last two possibilities give uninteresting models, at least from the point
of view of defects inside defects: the third possibility leads to a system of
two decoupled scalar fields and the fourth one severely restricts the number
of nontrivial vacua, forbidding the presence of defects inside defects.

\subsection{Topological solutions}

We consider the equations of motion (\ref{eqmx1}) and (\ref{eqmx2})
for the general model described by $H(\phi,\chi)$, given by eq.(\ref{H_gen}).
We have 
\begin{eqnarray} 
\label{d2phi-gen} 
\frac{d^2\phi}{dx^2}&=&[\lambda(\phi^2-a^2)+\mu\chi^2+2\nu\phi\chi] 
(2\lambda\phi+2\nu\chi)\nonumber\\ 
& &+[2\mu\phi\chi+\nu\phi^2+\sigma(\chi^2-b^2)](2\mu 
\chi+2\nu\phi)~,\\ 
\label{d2chi-gen} 
\frac{d^2\chi}{dx^2}&=&[\lambda(\phi^2-a^2)+\mu\chi^2+2\nu\phi\chi] 
(2\mu\chi+2\nu\phi)\nonumber\\ 
& &+[2\mu\phi\chi+\nu\phi^2+\sigma(\chi^2-b^2)](2\mu 
\phi+2\sigma\chi)~. 
\end{eqnarray}
To find defects inside defects we have to search for pairs of solutions  
like ($\phi$,0) and (0,$\chi$). For the first pair we set  
$\chi=0$ and then the above equations reduce to 
\begin{equation} 
\label{d2phi,chi=0} 
\frac{d^2\phi}{dx^2}=[2\lambda^2(\phi^2-a^2)+2\nu^2\phi^2- 
2\sigma\nu b^2]\phi~,
\end{equation}
and
\begin{equation} 
\label{d2chi,chi=0} 
\nu(\lambda+\mu)\phi^2-(\lambda\nu a^2+\mu\sigma b^2)=0~. 
\end{equation}
For the second pair we have $\phi=0$ and then Eqs.~(\ref{d2phi-gen})  
and (\ref{d2chi-gen}) read  
\begin{equation}
\label{d2chi,phi=0} 
\frac{d^2\chi}{dx^2}=[2\sigma^2(\chi^2-b^2)-2\mu\lambda a^2 
+2\mu^2\chi^2]\chi~, 
\end{equation}
and
\begin{equation}
\label{d2phi,phi=0} 
\mu(\nu+\sigma)\chi^2-(\lambda\nu a^2+\mu\sigma b^2)=0~. 
\end{equation}
Eqs.~(\ref{d2chi,chi=0}) and (\ref{d2phi,phi=0}) show that
solutions like defects inside defects are only possible for
\begin{equation} 
\label{cond123} 
\nu(\lambda+\mu)=0, \qquad \mu(\nu+\sigma)=0,  
\qquad \lambda\nu a^2+\mu\sigma b^2=0, 
\end{equation} 
Interestingly, these conditions are exactly the conditions
(\ref{cond1})-(\ref{cond3}), needed to ensure the $Z_2\times Z_2$ parity
symmetry of the system. This means that only the generating functions 
$H_1(\phi,\chi)$ and $H_2(\phi,\chi)$ may admit
solutions describing defects inside defects.

The explicit solutions can be found by solving the remaining equations. 
For the system described by $H_1$ we take $\nu=\sigma=0$ and
rewrite Eqs.~(\ref{d2phi,chi=0}) and (\ref{d2chi,phi=0}) as: $\chi=0$ and  
\begin{equation} 
\label{d2phi,H1} 
\frac{d^2\phi}{dx^2}= 2\lambda^2(\phi^2-a^2)\phi~,
\end{equation}
or $\phi=0$ and
\begin{equation}
\label{d2chi,H1} 
\frac{d^2\chi}{dx^2}= 2\mu^2(\chi^2- r a^2)\chi~, 
\end{equation} 
where we have set $\lambda/\mu=r$. There are pairs of solutions  
\begin{equation} 
\label{phi-kink} 
\phi(x)=a\tanh(\lambda ax)~; 
\qquad \chi=0~,
\end{equation} 
and
\begin{equation}
\label{chi-kink} 
\chi(x)=a\sqrt{r}\tanh(\mu\sqrt{r}ax)~; 
\qquad \phi=0~. 
\end{equation}
These solutions present the feature of vanishing at its own core, that is,
for $x\to0$ we get $\phi(x)\to0$ and $\chi(x)\to0$. This is important for
introducing defects inside defects in $3+1$ dimensions, since in this case
we can choose $\phi=\phi(x)$ and $\chi=\chi(y)$, say, to make topological
defect appear in the core of topological defect. The issue of choosing
the field to host the second field depends on the parameters of the system,
and follows as in the former investigation \cite{bba97} -- see also
Ref.~{\cite{etb98} for similar considerations in $2+1$ dimensions.  

Similarly, we can also find solutions like defects
inside defects in the other system, defined via $H_2(\phi,\chi)$. In 
this case the pairs of solutions are given by 
\begin{equation} 
\label{phi-kink2} 
\phi(x)=b\sqrt{\frac{\sigma}{\nu}} 
\tanh(\nu\sqrt{\frac{\sigma}{\nu}}bx)~; 
\qquad\chi=0~, 
\end{equation}
and 
\begin{equation}
\label{chi-kink2} 
\chi(x)=b\tanh(\sigma bx)~;\qquad\phi=0~. 
\end{equation} 
Also they may be used to build defects inside defects in the $3+1$
dimensional system.

\subsection{BPS solutions}
 
Let us now consider the first-order equations (\ref{foeq1}) and (\ref{foeq2})  
for the general $H$ given by Eq.~(\ref{H_gen}). Here we have 
\begin{eqnarray} 
\label{dphi_gen} 
\frac{d\phi}{dx}=\lambda(\phi^2-a^2)+\mu\chi^2+2\nu\phi\chi~,\\ 
\label{dchi_gen} 
\frac{d\chi}{dx}=\sigma(\chi^2-b^2)+\nu\phi^2+2\mu\phi\chi~. 
\end{eqnarray} 
We set $\chi=0$ to get to 
\begin{equation} 
\label{dphi,chi=0} 
\frac{d\phi}{dx}=\lambda(\phi^2-a^2)~,
\end{equation}
and
\begin{equation}
\label{dchi,chi=0}
\nu\phi^2-\sigma b^2=0~.
\end{equation}
We choose $\nu=\sigma=0$ to get to the pair of solutions
\begin{equation}
\label{phi-BPS-kink1}
\phi(x)=a\tanh \lambda ax~;\qquad\chi=0~.
\end{equation}
Analogously, in the case $\phi=0$ we have
\begin{equation}
\label{dchi,phi=0}
\frac{d\chi}{dx}=\sigma(\chi^2-b^2)~,
\end{equation}
and
\begin{equation}
\label{dphi,phi=0} 
\mu\chi^2-\lambda a^2=0 ~. 
\end{equation}
We impose $\lambda=\mu=0$ to obtain 
\begin{equation}
\label{phi-BPS-kink2} 
\chi(x)=b\tanh \sigma bx~;\qquad \phi=0~. 
\end{equation} 
These are BPS defects, since they arise from the first-order equations
(\ref{foeq1}) and (\ref{foeq2}).

It is interesting to realize that the above BPS solutions cannot be 
solutions of a single system, because the conditions
$\nu=\sigma=0$ and $\lambda=\mu=0$ cannot be implemented simultaneously in the
general $H$ given by Eq.~(\ref{H_gen}). This conclusion is in agreement with
results of dynamical systems, since one knows from unicity of solutions that
any two orbits never cross each other in configuration space. And we recall
that the pair of first-order equations (\ref{dphi_gen}) and (\ref{dchi_gen})
can be seen as a dynamical system. However, we can have the
BPS defect (\ref{phi-BPS-kink1}) in the system defined via $H_1(\phi,\chi)$,
and the BPS defect (\ref{phi-BPS-kink2}) in the other system, defined via
$H_2(\phi,\chi)$.

The above solutions have been investigated before \cite{bba97,mor98}, but here
we have a stronger result, showing that systems engendering solutions like
defects inside defects must present $Z_2\times Z_2$ symmetry. Furthermore,
only one of the solutions can be of the BPS type. Since the BPS solution is
stable \cite{baz95}, we still have to investigate stability of the other
solution to talk about stability of defects with internal structure.

For completeness let us investigate the presence of other BPS solutions
in the system defined by $H_1(\phi,\chi)$, for instance. In this case the
first-order equations are
\begin{eqnarray} 
\label{dphi_h1} 
\frac{d\phi}{dx}&=&\lambda(\phi^2-a^2)+\mu\chi^2~,\\ 
\label{dphi_h2} 
\frac{d\chi}{dx}&=&2\mu\phi\chi~. 
\end{eqnarray}
An interesting pair of BPS solutions was already obtained in \cite{baz95}.
It is
\begin{eqnarray}
\phi(x)&=&-a\tanh(2\mu ax)~,
\\
\chi(x)&=&\pm a\sqrt{ {r-2}\,} {\rm sech}(2\mu ax)~.
\end{eqnarray}
It is valid for $r=\lambda/\mu>2$ and obeys
\begin{equation}
\phi^2 + \frac{1}{\sqrt{r-2}}\,\chi^2=a^2~,
\end{equation}
which defines a semi-ellipsis in configuration space. We recall that at $r=3$
the amplitude of the two above solutions degenerate to a single value, the
orbit changes to a semi-circle and the corresponding stability can be
implemented analytically \cite{bnr97}.

This system presents peculiar behavior at two other values of $r$. For $r=1$
the symmetry changes from $Z_2\times Z_2$ to $Z_4$, but now the system
degenerates into two uncoupled systems of one field \cite{bba97}.
For $r=-1$ the vacuum states at the $\chi$ axis desapear,
and although there is still
the pair of BPS solutions given by Eq.~(\ref{phi-BPS-kink1}), it is possible
to show that solutions like the above one, describing a finite orbit connecting
the two vacuum states $(-a,0)$ and $(a,0)$ with non-vanishing $\chi$ cannot
be present anymore. The proof follows after recognizing that for $r=-1$
the first-order equations $(\ref{dphi_h1})$ and $(\ref{dphi_h2})$ implies that
\begin{equation}
\chi\,\left(\phi^2-\frac{1}{3}\chi^2-a^2\right)=0~.
\end{equation}
This restriction shows that $\chi=0$ or
\begin{equation}
\label{condr=-1}
\phi^2=a^2+\frac{1}{3}\,\chi^2
\end{equation}
The first condition $\chi=0$ gives exactly the case of the straight line
connecting the vacua $(-a,0)$ and $(a,0)$. The other condition
Eq.~(\ref{condr=-1}) shows that for $\chi\neq0$ no finite orbit can connect
the vacuun states anymore.

\subsection{Periodic solutions} 

Let us investigate the same systems given by $H_1$ and $H_2$, but now 
searching for periodic solutions of the second-order equations 
(\ref{eqmtx1}) and (\ref{eqmtx2}) that appear when the line becomes a circle, 
a compact space. In this case the equations of motion do not 
change but the boundary conditions are now periodic ones, to respect the 
topology of the circle. The solutions here are like the sphalerons 
solutions introduced in \cite{msa88} in the case of a single field -- 
see also \cite{lmk92,spha} and references therein for further 
details. For generality we shall assume that the fields $\phi$ and 
$\chi$ are periodic according to  
\begin{eqnarray} 
\phi(x)&=&\phi(x+L_1)~,\\
\chi(x)&=&\chi(x+L_2)~.
\end{eqnarray}
Evidently, when these fields are immersed in a one-dimensional space we 
should impose that $L_1=L_2$, but in higher dimensions we can search for 
solutions with $\phi=\phi(x)$ and $\chi=\chi(y)$, for instance, and so 
we can allow $L_1\ne L_2$ and this is the case we shall consider below. 
Here the solutions will be given in terms of the Jacobi elliptic 
function ${\rm sn} (x)$, which has period $4 K(k_i)$ determined by the 
elliptic quarter period $K(k_i)$ which obeys ($k_i\in (0,1)$, $i=1,2$) 
\cite{abramowitz}  
\begin{equation} 
\frac{1}{2}\pi=K(0)\le K(k_i)\le K(1)=\infty~. 
\end{equation} 
For the model $H_1$ we get the solutions: $\chi=0$ and 
\begin{equation}
\label{pp1}
\phi(x)=a k_1\, b(k_1)\;{\rm sn}\left(a\lambda b(k_1)\, x\right)~, 
\end{equation}
and also $\phi=0$ and 
\begin{equation}
\label{pp2}
\chi(x)=a k_2\sqrt{r}\,b(k_2)\;{\rm sn}\left(a\mu\sqrt{r}\,b(k_2)\; x\right)~, 
\end{equation} 
where
\begin{equation} 
b(k_i)=\sqrt{2\over 1+k^2_i},\hspace{1cm} i=1,2. 
\end{equation} 
The stability of these solutions will be studied in detail in the next 
section. Analogous solutions for $H_2$ can also be found by just changing
$\phi\to\chi$ and redefinig the parameters conveniently.

In the limit $k_i\to 1$ the radius of the circle goes to infinity, 
${\rm sn}(b(k_i) x)\to\tanh(x)$, and the above solutions recover 
the corresponding basic kink solutions we have already found 
in the previous subsections. Here we notice that in two or more space 
dimensions we may set $k_1\to1$ or $k_2\to1$, to make periodic or topological
defects to host topological or periodic defects, allowing for new pictures of
defects inside defects. As one can check, none of the above periodic solutions
of the second-order equations of motion solves the first-order equations,
despite the fact that the limit $k_i\to 1$ recover all the topological
solutions including the BPS ones.

\section{Linear stability} 
\label{sec:stability} 
 
In this section we investigate the stability of the solutions presented 
in the former Sec.~{\ref{sec:solutions}}. Evidently, a general
study of the stability of static solutions of nonlinear 
equations is a highly nontrivial matter, but the simplest situation 
known as classical or linear stability may be performed and will be 
analysed with some detail below. Firstly we recall that the solutions 
(\ref{phi-kink}) and (\ref{chi-kink2}) also solve the first-order 
equations, and so are linearly stable \cite{baz95}. All other solutions
should be investigated with respect to their stability.

We start by writing the time-dependent 
fields in the usual form \cite{jac77,raj82} 
\begin{eqnarray} 
\phi(x,t)&=&\phi(x)+\eta(x,t)~,\\ 
\chi(x,t)&=&\chi(x)+\zeta(x,t)~, 
\end{eqnarray} 
where $\eta(x,t)$ and $\zeta(x,t)$ are the time-dependent perturbations,
small when compared to $\phi(x)$ and $\chi(x)$, the static solutions of the
second-order equations of motion. Since we are considering fluctuations about
static solutions, we can always write 
\begin{eqnarray} 
\eta(x,t)&=&\sum_n \eta_n(x)\cos(w_n t)~,\\ 
\zeta(x,t)&=&\sum_n \zeta_n(x)\cos(w_n t)~. 
\end{eqnarray} 
Substituting these expressions into the time-dependent Euler-Lagrange 
equations of motion we get the following Schr\"odinger equation 
\begin{equation} 
\label{me} 
\left(- {\bf{1}} \frac{d^2}{dx^2} + {\bf M} \right) 
\left({\eta_n \atop \zeta_n}\right) 
= w^2_n \left({\eta_n \atop \zeta_n}\right)~, 
\end{equation} 
where ${\bf{1}}$ is the $2\times 2$ identity matrix and 
\begin{equation}\label{M} 
{\bf M} = 
\left({U_{\phi\phi} \;\;\;\;\; U_{\phi\chi} 
\atop U_{\chi\phi} \;\;\;\;\; U_{\chi\chi}}\right)~. 
\end{equation} 
The derivatives of the potential $U(\phi,\chi)$ should be 
calculated at the classical values $\phi(x)$ and $\chi(x)$, and we 
recall that stability means absence of negative $w^2_n$. 

The standard analysis of stability usually considers diagonalizing the above
matrix ${\bf M}$, but this may make the problem untractable analytically.
We illustrate this situation by following the lines of Ref.~{\cite{bnr97}},
where instead of considering the matrix involving derivatives of the potential
$U(\phi,\chi)$, one considers another one, involving derivatives of
$H(\phi,\chi)$. This procedure follows naturally from the fact that the
first order equations can be used to simplify investigations concerning the
second-order equations, at least when the classical configurations also solve
the first-order differential equations. For BPS solutions this is implemented
after introducing the first-order operators
\begin{equation}
\label{ope1}
S_\pm=\pm{\bf 1}\frac{d}{dx}+{\bf m}~,
\end{equation}
where {\bf m} is the matrix
\begin{equation}
{\bf m}=\left(
\begin{array}{cc}
H_{\phi\phi} & H_{\phi\chi}\\
H_{\chi\phi} & H_{\chi\chi}
\end{array}
\right)~,
\end{equation}
so that the stability equation (\ref{me}) can be rewritten as
\begin{eqnarray}
\label{eqsta}
\left[ -{\bf 1}\frac{d^2}{dx^2}+\frac{d{\bf m}}{dx} +{\bf m}^2\right]
\left({\eta_n \atop \zeta_n}\right)&=&
S_+S_- \left({\eta_n \atop \zeta_n}\right)\\
&=& w^2_n\,\left({\eta_n \atop \zeta_n}\right)~.
\end{eqnarray}
In (\ref{eqsta}) the Schr\"odinger-like operator presents the form
$S_{+}S_{-}$. However, in Eq.~(\ref{ope1}) the first-order operators are
such that $S^{\dag}_{\pm}=S_{\mp}$. Thus we can write
\begin{equation}
w^2_n=<n|S_+S_-|n>=<n|S^{\dag}_-S_-|n>=||S_-|n>||^2\geq0~,
\end{equation}
where $|n>$ stands for the (orthonormalized) state
\begin{equation}
|n>=\left({\eta_n \atop \zeta_n}\right)~.
\end{equation}
This shows explicitly that the Schr\"odinger-like operator in
Eq.~(\ref{eqsta}) is positive semi-definite, and so the corresponding
eigenvalues $w^2_n$ are non-negative -- see Ref.~{\cite{baz95}}
for further details. The BPS solutions are stable,
but if we want to find the eigenvalues explicitly we need to diagonalize
the corresponding potential. This diagonalization for BPS solutions can be
given in terms of the diagonalization of the simpler matrix {\bf m}.
In this case the elements of the diagonalized matrix are
\begin{equation}
\label{mdiag}
\lambda_{\pm}=\frac{1}{2}(H_{\phi\phi}+H_{\chi\chi})
\pm\frac{1}{2}\sqrt{(H_{\phi\phi}-H_{\chi\chi})^2+4 {H_{\phi\chi}}^2}~,
\end{equation}
and the corresponding Schr\"odinger-like equations are given by
\begin{equation}
\label{Se1}
\left(-\frac{d^2}{dx^2} + \frac{d\lambda_\pm}{dx}
+(\lambda_\pm)^2\, \right) \eta_n^{\pm}= w^2_n\, \eta_n^{\pm}~.
\end{equation}

The problem of solving the above equations analytically is now related to
the presence of the square root in the potential of the corresponding
Schr\"odinger equations. To mantain the investigation analytical,
we should get rid of the square root in Eq.~(\ref{mdiag}), and this can
be done by imposing one of the following conditions
\begin{eqnarray}
\label{rc1}
&{\rm (i):}&\, H_{\phi\chi}=0 ~,\\
\label{rc2}
&{\rm (ii):}&\, H_{\phi\phi}=H_{\chi\chi}~,\\
\label{rc3}
&{\rm (iii):}&\, H_{\phi\phi} H_{\chi\chi}={H_{\phi\chi}}^2~.
\end{eqnarray}
For the general model, the above condition (i)
can be satisfied with the following possibilities
\begin{eqnarray}
&(\rm{ia}):&\,\mu=\nu=0 ~,\nonumber\\
&(\rm{ib}):&\,\nu=0\,\,{\rm and}\,\,\chi=0 ~,\nonumber\\
&(\rm{ic}):&\,\mu=0\,\,{\rm and}\,\, \phi=0 ~.\nonumber
\end{eqnarray}
The possibility (ia) must be discarded since it decouples the two fields. The
other two possibilities (ib) and (ic) can be considered and here $\phi$ in
case (ib) and $\chi$ in case (ic) may describe topological domain defects
of the BPS type, but in these cases no defect is allowed to appear inside it.

The second possibility of eliminating the square root from the stability
equation is $H_{\phi\phi}=H_{\chi\chi}$. Since the generalized model presents
$H_{\phi\phi}=2\lambda\phi+2\nu\chi$ and $H_{\chi\chi}=2\mu\phi+2\sigma\chi$
we see that the condition (ii) is obtained by imposing $\lambda = \mu$ and
$\sigma = \nu$. In  this case we have
\begin{equation}
H_s=\mu\left(\frac{1}{3}\phi^3-a^2\phi\right)+ \mu\phi\chi^2+\nu\phi^2\chi+
    \nu\left(\frac{1}{3}\chi^3-b^2\chi\right)~.
\end{equation}
This function can be written in the form $H_s={\bar H}_1+{\bar H}_2$, where
\begin{eqnarray}
{\bar H}_1&=&\mu\left(\frac{1}{3}\phi^3-a^2\phi\right)+ \mu\phi\chi^2~,
\\
{\bar H}_2&=&\nu\left(\frac{1}{3}\chi^3-b^2\chi\right)+ \nu\chi\phi^2~.
\end{eqnarray}
In this case, however, we follow \cite{bba97} to see that each
${\bar H}_i\,(i=1,2)$ can be written in terms of
$\phi_{\pm}=2^{-1/2}\,(\phi\pm\chi)$, in a form that decouple the two fields,
and so also does the full system described by $H_s$. Thus we see that
the general model does not give any system of two coupled fields
under the additional condition $H_{\phi\phi}=H_{\chi\chi}$.

The third and last condition (iii) implies
\begin{equation}
\lambda\mu\phi^2+\sigma\nu\chi^2+(\lambda\sigma+\mu\nu)\phi\chi=
\nu^2\phi^2+\mu^2\chi^2+2\mu\nu\phi\chi~,
\end{equation}
which is satisfied when we identify $\lambda\mu=\nu^2$ and
$\sigma\nu=\mu^2$, which also implies $\lambda\sigma=\mu\nu$. These
conditions lead to a simpler system, but such system describes no system
of two coupled fields anymore.

We see that none of the conditions (\ref{rc1})-(\ref{rc3}) leads to
systems of coupled fields. However, Eq.~(\ref{mdiag})
shows another possibility, which appears as follows: we investigate
the quantity
\begin{equation}
\left(H_{\phi\phi}-H_{\chi\chi}\right)^2+4\,H_{\phi\chi}
\end{equation}
with the aim of eliminating the square root in Eq.~(\ref{mdiag}),
restricting the set of parameters that defines $H(\phi,\chi)$.
This possibility is particular, and can only be implemented in specific 
systems, where $H(\phi,\chi)$ is explicitly known. For instance,
in Ref.~{\cite{bnr97}} one uses this reasoning to present analytical
calculations that allow obtaining the discrete eigenvalues corresponding
to BPS solutions of systems like the ones described by (\ref{H_1})
or (\ref{H_2}), in a specific region of the space of parameters.

The above results show that although the BPS solutions are stable, in general
it is not ease to know the corresponding eigenvalues explicitly. Evidently,
we can always give up the analytical procedure to solve Schr\"odinger-like
equations like Eq.~(\ref{Se1}) numerically, but this is out of the scope
of the present work. From the point of view of defects inside
defects, the important information is that the BPS solutions are stable
configurations, and so they can be considered stable defects to host
non-BPS defects. The stability of the non-BPS defects will be investigated
in the next subsection.

The result that the general system does not give any system of two coupled
fields for $H_{\phi\phi}=H_{\chi\chi}$ can be
generalized in the following way. We introduce the new fields
$\phi_{\pm}=2^{-1/2}\,(\phi\pm\chi)$ and use the equations of motion
(\ref{eqmx1}) and (\ref{eqmx2}) to obtain
\begin{eqnarray}
\frac{d^2\phi_{+}}{dx^2}&=&H_{+}H_{++} + H_{-}H_{-+}~,
\\
\frac{d^2\phi_{-}}{dx^2}&=&H_{+}H_{+-} + H_{-}H_{--}~.
\end{eqnarray}
Here we are using the notation $H_{+}=\partial H/\partial\phi_{+}$,
$H_{-}=\partial H/\partial\phi_{-}$, and so forth. Derivatives of the
function $H$ are related by
\begin{eqnarray}
H_{\phi}&=&\frac{1}{\sqrt{2}}\,(H_{+}+H_{-})~,
\\
H_{\chi}&=&\frac{1}{\sqrt{2}}\,(H_{+}-H_{-})~,
\end{eqnarray}
and also
\begin{eqnarray}
\label{h11}
H_{\phi\phi}&=&\frac{1}{2}\,H_{++}+\frac{1}{2}\,H_{--}+
\frac{1}{2}\,H_{+-}+\frac{1}{2}\,H_{-+}~,
\\
\label{h22}
H_{\chi\chi}&=&\frac{1}{2}\,H_{++}+\frac{1}{2}\,H_{--}-
\frac{1}{2}\,H_{+-}-\frac{1}{2}\,H_{-+}~,
\\
\label{h12}
H_{\phi\chi}&=&\frac{1}{2}\,H_{++}-\frac{1}{2}\,H_{--}+
\frac{1}{2}\,H_{-+}-\frac{1}{2}\,H_{+-}~,
\\
\label{h21}
H_{\chi\phi}&=&\frac{1}{2}\,H_{++}-\frac{1}{2}\,H_{--}+
\frac{1}{2}\,H_{+-}-\frac{1}{2}\,H_{-+}~.
\end{eqnarray}
We use (\ref{h12}) and (\ref{h21}) to see that the condition
$H_{\phi\chi}=H_{\chi\phi}$ now becomes $H_{+-}=H_{-+}$. Furthermore,
from (\ref{h11}) and (\ref{h22}) we realize that if
one further imposes the condition $H_{\phi\phi}=H_{\chi\chi}$ one gets
$H_{+-}=H_{-+}=0$. This result is general, and implies that
$H(\phi_{+},\phi_{-})$ can be written as the sum of two functions, one
depending on $\phi_{+}$ and the other on $\phi_{-}$. The Lagrangian density
is then reduced to a sum of two Lagrangian densities, one for the field
$\phi_{+}$ and the other for $\phi_{-}$. The system decouples into two
systems of a single field each one, and so it does not describe two coupled
fields anymore.

\subsection{Stability of the topological solutions} 
 
The pair of solutions with $\phi=0$ and $\chi$ given by 
Eq.~$(\ref{chi-kink})$ constitutes a pair of non-BPS solutions, that is, 
it does not obey the corresponding first-order equations. Then its 
stability should be investigated explicitly, and we do it now since it 
was not done in the former works \cite{brs96,bba97}. Here, we have that 
$U_{\phi\chi}$ vanishes at the classical values with $\phi=0$, and we 
are left with the following uncoupled Schr\"odinger-like equations, 
after appropriately rescaling the space coordinate, 
 
\begin{eqnarray} 
\left(\frac{d^2}{dz^2}+\frac{w_n^2}{r\mu^2a^2}-4+2(2+r)\, 
{\rm sech}^2 z \right) \eta_n(z)&=&0~,\\ 
\left(\frac{d^2}{dz^2}+\frac{w_m^2}{r\mu^2a^2}-4+6\, 
{\rm sech}^2 z \right) \zeta_m(z)&=&0~, 
\end{eqnarray} 
which are well known modified Posch-Teller equations \cite{mfe53} 
whose eigenvalues are given by 
\begin{eqnarray} 
\frac{w^2_{n}}{r\mu^2a^2}&=&4-\Biggl[\sqrt{2(2+r)+\frac{1}{4}} 
-\left(n+\frac{1}{2}\right)\,\Biggr]^2~,\\ 
\frac{w^2_{m}}{r\mu^2a^2}&=&4-\Biggl[\frac{5}{2} 
-\left(m+\frac{1}{2}\right)\,\Biggr]^2~, 
\end{eqnarray} 
where $n=0,1,..,<\sqrt{2(2+r)+1/4}-1/2$ and $m=0,1$. 
Since $r>0$ we find the following restriction on the solution (\ref{chi-kink}) 
of the parity preserving model $H_1$ 
\begin{equation} 
0 < r\equiv\frac{\lambda}{\mu} \le 1~. 
\end{equation} 
However, since we assume that $r\ne 1$, in order to keep the fields 
coupled, we have $r\in(0,1)$ as the range of values that ensure 
stability of the non-BPS solution $(\ref{chi-kink})$. Similar conclusions
can be found for the other system, with $H_2$. 
 
\subsection{Stability of the periodic solutions}

We now discuss the stability of the periodic solutions (\ref{pp1}) and 
(\ref{pp2}). The general discussion on linear stability already 
presented applies equally well here, the difference in the present case 
is that the equations that appear for the fluctuations are now of the 
Lam\'e type 
 
\begin{equation} 
\left\{{d^2\over dz^2}+h-N(N+1){\rm sn}^2(z)\right\}\,f(z)=0~. 
\end{equation} 
As we are going to show, some of these equations are solved by the 
Lam\'e polynomials, since $N$ can be identified with integer. But there 
are others for which $N$ is not integer and so the exact solutions will 
be not completely known \cite{ars64}. 
Evidently, for each solution we have two equations of stability, the first 
describing fluctuation in $\phi$ and the second in $\chi$. 
For the periodic solution $(\ref{pp1})$ the stability equations are 
\begin{eqnarray} 
\left\{{d^2\over dz^2}+(1+k^2)\left({\omega^2\over 2a^2\mu^2r^2}+1\right) 
-6k^2{\rm sn}^2(z)\right\}\,\eta(z)=0~, 
\\ 
\left\{{d^2\over dz^2}+(1+k^2)\left({\omega^2\over 2a^2\mu^2r^2}+{1\over r} 
\right)-{2\over r}(1+{2\over r})k^2{\rm sn}^2(z)\right\}\,\zeta(z)=0~, 
\end{eqnarray} 
where we omit indices on eigenfunctions and eigenvalues and also on 
$k_1$ and $k_2$, for simplicity. The first stability equation is of the 
Lam\'e type with $N=2$, so that there are $2N+1=5$ solutions whose 
eigenvalues are 
\begin{equation} 
\label{w2} 
\frac{{w}^2}{r^2 a^2\mu^2 b^2(k)}=3,\,3k^2,\,0,\,1 
+k^2\pm 2\sqrt{1-k^2+k^4}~. 
\end{equation} 
Here, instability appears because of the last eigenvalue with the minus sign, 
which is always negative for any $k$ since $0\le k^2 <1$.   
 
The second stability equation is also of Lam\'e 
type but now $N=2/r$ is not an integer, in general. The solutions for this 
case are not completely known but are expressible as series of 
elliptic functions which truncate to a polynomial when $N$ becomes an 
integer. Also, the corresponding eigenvalues for $h$ are not known in 
general, except for a few cases like $N=1/2$ or $3/2$. So, for special 
choices of the ratio $r$ 
\begin{center} 
\begin{tabular}{c| c c c c c c} 
$r$ & 4   & 2 & 4/3 & 1 & 2/3 & \dots\\ \hline 
$N$ & 1/2 & 1 & 3/2 & 2 & 3   & \dots\\ 
\end{tabular} 
\end{center} 
we are able to find explicit solutions and the corresponding eigenvalues. 
Let us analyse some of these possibilities. For $r=2$, $N=1$  the 
eigenvalues are: 
\begin{equation} 
{\omega^2\over 4a^2\mu^2 b^2(k)}=1,\,{1-k\over 1+k},\, 
{k-1\over 1+k}~, 
\end{equation} 
which are non-negative except for the last one. For the case $r=1$, $N=2$ 
the eigenvalues are 
\begin{equation} 
\frac{\omega^2}{a^2\mu^2b^2(k)}=3,3k^2,0, 
1+k^2\pm 2\sqrt{1-k^2+k^4}~, 
\end{equation} 
which are positive except for the last one with the minus sign, which is 
negative for any $k\in(0,1)$. Let us also comment on one case where $N$ 
in half-integer: we choose it to be $N=1/2$, which corresponds to $r=4$; 
in this case the eigenvalue is zero, exactly. 

A similar situation appears for the study of the stability of the second 
periodic solution $(\ref{pp2})$, for which the stability equations are 
\begin{equation} 
\left\{{d^2\over dz^2}+(1+k^2)\left({\omega^2\over 2a^2\mu^2r}+1\right) 
-6k^2{\rm sn}^2(z)\right\}\,\zeta(z)=0~, 
\end{equation} 
\begin{equation} 
\left\{{d^2\over dz^2}+(1+k^2)\left({\omega^2\over 2a^2\mu^2r} 
+{1\over r}\right) 
-2(2+r)k^2{\rm sn}^2(z)\right\}\,\eta(z)=0~. 
\end{equation} 
Here we also find that $N=2$ for the first stability equation, 
and in this case the eigenvalues are 
\begin{equation}\label{w3} 
\frac{\omega^2}{a^2\mu^2 r b^2(k)k^2}=3,\,3k^2,\,0,\, 
1+k^2\pm 2\sqrt{1-k^2+k^4}~. 
\end{equation} 
Analogously to the previous case, the second stability equation for  
the solution $(\ref{pp2})$ implies 
\begin{equation}
N=-\frac{1}{2}+\frac{1}{2}\sqrt{1+8(2+r)}
\end{equation}
 so that in general we will 
have no closed solution except for special values of the ratio $r$ 
\begin{center} 
\begin{tabular}{c| c c c c c c} 
$r$ & 1 & 4 & 8 & 13 &\dots\\ \hline 
$N$ & 2 & 3 & 4 & 5  &\dots\\ 
\end{tabular} 
\end{center} 
Note that lower values of $N$ are not allowed since we imposed $r>0$. 
However, if we choose $r=1$, $N=2$ we have the eigenvalues 
\begin{equation}
\frac{\omega^2}{a^2\mu^2b^2(k)}=3,\,3k^2,\,0,\,
1+k^2\pm 2\sqrt{1-k^2+k^4}~,
\end{equation}
and the periodic solution (\ref{pp2}) is also unstable, as expected. 

We notice that the above stability equations 
recover the corresponding stability equations of the former subsection 
in the limit $k\rightarrow 1$. Since sn$^2 z\rightarrow \tanh^2 z = 1 
-$sech$^2 z$, we see that the operator 
\begin{equation} 
\frac{d^2}{dz^2}+h_k-k^2p\,{\rm sn}^2z~, 
\end{equation} 
changes to 
\begin{equation} 
\frac{d^2}{dz^2}+(h_1-p)+p\,{\rm sech}^2 z~.
\end{equation} 
However, there are subtleties in the limit $k\to1$ for the eigenfunctions and 
eigenvalues of the corresponding periodic and topological solutions. For
instance, it is only when $k=1$ that we have BPS solutions; for $k\neq1$ the
periodic solutions will never solve the first-order equations. See
Refs.~{\cite{msa88,lmk92}} for more detais on this issue in the case of a
single scalar field.

\subsection{Energy and stability} 

In this subsection we introduce another reasoning, that leads to 
another condition to be fulfilled by the classical solutions 
in order to make them stable against decaying into less energetic solutions. 
This reasoning is simple and can be interesting, mainly when one is 
unable to implement the standard investigation of stability analytically. 
The reasoning is based on the presence of stable BPS solutions, which solve 
the first-order equations, and so it does not work for periodic solutions.

Let us consider the parity preserving model defined with $H_1$. In this 
case the non-BPS solution obtained with $\phi=0$ has energy $E=(4/3)\mu 
r^{3/2}a^3$, and connects the two vacuum states $(0,a\sqrt{r})$ and 
$(0,-a\sqrt{r})$. However, these two vacuum states can be also connected 
by considering BPS solutions that make use of the vacuum $(a,0)$ or 
$(-a,0)$. Evidently, there are two degenerate possibilities of 
connecting the vacua $(0,a\sqrt{r})$ and $(0,-a\sqrt{r})$: one uses
$(-a,0)$ and the other $(a,0)$, choosing the left and right path, respectively.
The energy of these two BPS solutions can be calculated easily -- see
Sec.~{\ref{sec:intro}}. The result is $(4/3)\mu r a^3$. We compare this
with the energy $E=(4/3)\mu r^{3/2}a^3$ of the non-BPS solution to see that
it is only for $r\le1$ that the non-BPS defect does not decay into a pair of
BPS defects. Interestingly, this is the same result we have already obtained
using the standard investigation of classical or linear stability.

\section{High temperature effects}
\label{sec:temperature}
 
Owing to the possibility of applications to cosmology, let us now
study the effective potential at finite temperature for the models introduced
in this work. The effective potential allows introducing the conditions
for symmetry restoration, to identify the symmetric phase, the phase in
which the system supports no topological defects anymore. Here the
investigations are done in the (3+1) dimensional space-time, and deal
with constant and uniform field configurations. The one-loop effective
potential and the corresponding finite temperature effects can be
calculated according to the standard investigations \cite{temp}.
It can be written as, keeping only the high temperature
contributions \cite{bba97}, 
\begin{equation} 
\label{U^T} 
U^T(\phi,\chi)=U(\phi,\chi)+\frac{T^2}{24}(U_{\phi\phi}+U_{\chi\chi})~. 
\end{equation} 
The above expression for the effective potential shows that 
the thermal corrections add to mass terms in the following form 
\begin{eqnarray} 
{\sl m}^2_{\phi\phi}(T) &=& U_{\phi\phi}(0,0)\label{mphiphiT} 
        +\frac{T^2}{24}(U_{\phi\phi\phi\phi}+U_{\chi\chi\phi\phi})~,\\  
{\sl m}^2_{\chi\chi}(T) &=& U_{\chi\chi}(0,0)\label{mchichiT} 
        +\frac{T^2}{24}(U_{\phi\phi\chi\chi}+U_{\chi\chi\chi\chi})~,\\  
{\sl m}^2_{\phi\chi}(T) &=& U_{\phi\chi}(0,0)\label{mphichiT} 
        +\frac{T^2}{24}(U_{\phi\phi\phi\chi}+U_{\chi\chi\chi\phi})~. 
\end{eqnarray}
It is convenient to investigate the general model, defined with the potential
given by Eq.~(\ref{U_gen}). The effective 
potential at finite temperature in this case can be written as
\begin{eqnarray} 
\label{U^T_g} 
U^T(\phi,\chi) 
&=&\frac{1}{2}(\lambda^2+\nu^2)\phi^4 
        +2\nu(\lambda+\mu)\phi^3\chi 
        +(\lambda\mu+2\nu^2+\sigma\nu+2\mu^2)\phi^2\chi^2\nonumber\\ 
& &     +2\mu(\nu+\sigma)\phi\chi^3 
        +\frac{1}{2}(\mu^2+\sigma^2)\chi^4       
        +\frac{1}{2}{\sl m}^2_{\phi\phi} (T)\phi^2  
        +\frac{1}{2}{\sl m}^2_{\chi\chi} (T)\chi^2\nonumber\\ 
& &     +{\sl m}^2_{\phi \chi}(T)\phi\chi        
        +\frac{T^2}{24}[{\sl m}^2_{\phi\phi}(0)  
        +{\sl m}^2_{\chi\chi} (0)] 
        +\frac{1}{2}(\lambda^2a^4+\sigma^2b^4)~, 
\end{eqnarray} 
where the mass parameters at finite and zero temperature are given by 
\begin{eqnarray} 
\label{m_gphiphiT} 
{\sl m}^2_{\phi\phi} (T)&=&{\sl m}^2_{\phi\phi}(0) 
+\frac{T^2}{6}(3\lambda^2+5\nu^2+\lambda\mu+\sigma\nu+2\mu^2)~, 
\\ \label{m_gchichiT} 
{\sl m}^2_{\chi\chi} (T)&=&{\sl m}^2_{\chi\chi}(0) 
+\frac{T^2}{6}(3\sigma^2+\lambda\mu+2\nu^2+\sigma\nu+5\mu^2)~, 
\\ \label{m_gphichiT} 
{\sl m}^2_{\phi\chi} (T)&=&{\sl m}^2_{\phi\chi}(0) 
+\frac{T^2}{2}(\lambda\nu+\sigma\mu+2\mu\nu)~, 
\\ 
\label{mphiphi0} 
{\sl m}^2_{\phi\phi} (0)&=&-2(\lambda^2a^2+\sigma b^2\nu)~,\\ 
\label{mchichi0} 
{\sl m}^2_{\chi\chi} (0)&=&-2(\lambda a^2\mu+\sigma^2 b^2)~,\\ 
\label{mphichi0} 
{\sl m}^2_{\phi\chi} (0)&=&-2(\lambda a^2\nu+\sigma b^2\mu)~. 
\end{eqnarray}
In order to ensure that the potential mantains the 
$Z_2\times Z_2$ parity symmetry at finite temperature we impose, besides
the restrictions already found at zero temperature,
\begin{equation}
\lambda\nu +\sigma\mu +2\mu\nu=0. \label{cond4T} 
\end{equation} 
We notice that this condition is fully satisfied by the choices $\nu=\sigma=0$
or $\lambda=\mu=0$ corresponding to $H_1$ or $H_2$, respectively. This result
indicates that the existing solutions of second-order equations at 
zero temperature may persist when the system is in equilibrium with a 
thermal bath. However, we have been unable to write the finite temperature 
effective potential $U^T(\phi,\chi)$ via some generalized function 
$H^T(\phi,\chi)$ in the form $(\ref{potential})$ that appears at zero 
temperature. Although we have no explicit proof, we argue that the thermal
effects destroy the Bogomol'nyi bound, together with the possibility of
finding BPS defects at finite temperature. 
 
To define the masses properly and to introduce the critical temperatures 
we rotate the plane $(\phi,\chi)$ to the plane $(\phi_+,\phi_-)$ that 
diagonalizes the mass matrix. In this case we have ${\sl m}^2_{+}(T)>0$ 
and ${\sl m}^2_{-}(T)>0$ as the conditions for symmetry restoration in 
each one of the two independet field directions $\phi_+$ and $\phi_-$, 
respectively. The critical temperatures are obtained in the limit where 
these masses vanish, and here they are given by
\begin{equation} 
\label{e23} 
(T^c_\pm)^2=\frac{12}{\Lambda_1\Lambda_2-\Lambda_3^2} 
\Biggl[-(m_1\Lambda_2+m_2\Lambda_1-2m_3\Lambda_3)\pm\sqrt{\Delta}\Biggr]~, 
\end{equation} 
where we have set 
\begin{equation} 
\Delta=(m_1\Lambda_2-m_2\Lambda_1)^2+4m_1m_2\Lambda_3^2 
-4m_3(m_1\Lambda_2\Lambda_3+m_2\Lambda_1\Lambda_3-m_3\Lambda_1\Lambda_2)~. 
\end{equation} 
The parameters $m_i$ and $\Lambda_i$ are derivatives of 
$U(\phi,\chi)$ evaluated at the point $(0,0)$, and are given by 
\begin{eqnarray} 
\label{ee23} 
m_1&=&U_{\phi\phi}~,\qquad m_2=U_{\chi\chi}~,\qquad m_3=U_{\phi\chi}~, 
\nonumber\\ 
\Lambda_1&=&U_{\phi\phi\phi\phi}+U_{\phi\phi\chi\chi},\quad 
\Lambda_2=U_{\chi\chi\chi\chi}+U_{\phi\phi\chi\chi},\quad 
\Lambda_3=U_{\phi\phi\phi\chi}+U_{\chi\chi\chi\phi}. 
\end{eqnarray} 
 
Let us consider for instance a model defined by the general potential
$U(\phi,\chi)$ given by eq. (\ref{U_gen}) but with the restrictions 
$b^2=ra^2$, $\nu=r\sigma$, and $\lambda=r\mu$, so that its vacuum states are
$(\pm a,0)$ and $(0,\pm a\sqrt{r})$. The critical temperatures in this case
are given by 
\begin{eqnarray} 
\label{ee14} 
(T^c_\pm)^2&=&\frac{12a^2r(r+1)[(r+2)\sigma^2-(2r+1)\mu^2]} 
{(5r+1)(2r^2+r+3)\sigma^2-(r+5)(3r^2+r+2)\mu^2} 
\nonumber\\ 
& &\pm\frac{12a^2r(r-1)\sqrt{(r-1)^2(\sigma^4+\mu^4) 
+(6r^2+4r+6)\mu^2\sigma^2}} 
{(5r+1)(2r^2+r+3)\sigma^2-(r+5)(3r^2+r+2)\mu^2}~. 
\end{eqnarray} 
Then, to find the critical temperatures for the parity preserving model 
$H_1$ we just take $\sigma=0$ on the above result so that we have 
\begin{eqnarray} 
(T^c_+)^2&=&\frac{12r^2a^2}{3r^2+r+2}~, 
\\ 
(T^c_-)^2&=&\frac{12ra^2}{r+5}~. 
\end{eqnarray} 
This result coincide with the one recently 
found in Ref.~\cite{bba97}.
 
It is also interesting to compute the critical temperatures for the 
model defined by $H_2$, Eq.~(\ref{H_2}). They are also obtained from
Eq.~(\ref{ee14}), but now imposing $\mu=0$. Here we get 
\begin{eqnarray} 
(T^c_+)^2&=&\frac{12ra^2}{5r+1}~, 
\\ 
(T^c_-)^2&=&\frac{12ra^2}{2r^2+r+3}~. 
\end{eqnarray}  
We notice  that these results for $H_2$ are compatible with the  
ones obtained for $H_1$ since they are connected via the identifications: 
$ra^2\leftrightarrow a^2$ and $r\leftrightarrow 1/r$. 

The existence of two critical temperatures is due to the fact 
that the systems we are studing have two degrees of freedom and as a 
consequence their symmetries are restored separately for each one of the 
two independent field directions.
 
\section{Comments and conclusions} 
\label{sec:comments} 
 
In this paper we have dealt with systems of two coupled real scalar 
fields, searching for and investigating the corresponding classical or 
linear stability of the basic topological and periodic solutions when one 
of the two fields is set to zero. Our investigation is related to the
possibility of nesting defects inside defects in $3+1$ spacetime dimensions.
However, in $3+1$ dimensions renormalization restricts interactions to the
fourth power in the polynomial potential, and so we have been mostly concerned
with kinks of the hyperbolic tangent type. Such kinks have the property of
vanishing at its own core or center, and so at the core of the host defect
the system is reduced to a system of a single field, which is still able to
generate the other defect, the defect to be nested inside the host defect.

We have shown that the $Z_2\times Z_2$ parity symmetry is necessary
to the formation of defects inside defects of this kind. As we have seen, the 
two basic defects that appears when one of the two fields vanishes can 
be summarized as follows: 
\begin{center} 
\begin{tabular}{||c|c|c|c||}\hline\hline 
       &          & topological & nontopological \\  
 model & symmetry & solutions & solutions      \\   
\hline\hline 
$H_1$ & $Z_2\times Z_2$ &  $(\tanh x,0)$/BPS &  (sn $x$, 0)  \\ 
                &      & $(0,\tanh x)$/non-BPS & (0, sn $x$) \\  
\hline 
$H_2$ & $Z_2\times Z_2$ & $(\tanh x,0)$/non-BPS & (sn $x$, 0) \\ 
                    &       & $(0,\tanh x)$/BPS & (0, sn $x$) \\  
\hline 
$H_1$ or $H_2$  &   $ Z_4$  & $(\tanh x,0)$/BPS & (sn $x$, 0)\\ 
 $(r=1)$     &     & $(0,\tanh x)$/BPS & (0, sn $x$) \\  
\hline\hline 
\end{tabular} 
\end{center} 
We recall that in the last case, for $r=1$ the two $Z_2\times Z_2$ models 
colapse to the same $Z_4$ model, but this model does not describe a 
system of two {\sl coupled} fields anymore \cite{bba97}. 
 
We have also investigated the presence of periodic solutions when one of 
the two fields is set to zero. These solutions are direct extensions of 
the sphalerons found in \cite{msa88} in the case of a single real 
scalar field. The basic motivation for doing this is to enlarge the 
scope of the paper, since now one can mix topological and periodic 
defects, for instance allowing sphalerons to be nested inside domain 
walls. Furthermore, since the systems we have considered can be seen as real
bosonic portions of supersymmetric models, it seems also interesting to
investigate the behavior of fermions in such models.

The high temperature results show the presence of two critical temperatures,
signalling symmetry restoration in each one of the two independent field
directions. Here the thermal effects seems to destroy the Bogomol'nyi bound,
making the BPS defect to vanish at finite temperature. Within the standard
cosmological evolution the temperature decreases until breaking the symmetry
for the first field, allowing the formation of the host defect and later,
after the breakdown of the symmetry corresponding to the second field, the
second defect appears nested inside the first defect. As we can see, it is
possible to built models in which topological defects can nest nontopological
sphalerons, introducing instability inside the planar region on the wall,
which may be of interest to applications to cosmology. Other interesting
directions consider systems of coupled fields to investigate new aspects
of inflation \cite{kls94,cle96,bvh97,kls97} and the possibility of having
topological defects ending on topological defects \cite{ctr98,chw98}.
These and other related issues are presently under consideration. 

\begin{center} 
Acknowledgments 
\end{center} 

DB thanks J. R. Morris for interesting comments. DB and HBF thank the Center 
for Theoretical Physics, Massachusetts Institute of Technology for 
hospitality, and Conselho Nacional de Desenvolvimento Cient\'\i fico e 
Tecnol\'ogico, CNPq, Brazil, for partial support. FAB thanks F. Moraes and 
R. F. Ribeiro for discussions, and Coordena\c c\~ao de Apoio ao Pessoal do 
Ensino Superior, CAPES, Brazil, for a fellowship.

\end{document}